# Realization of a photonic topological insulator in Kagome crystals at terahertz wavelengths

Yun Shen[1, 2, 3*], Jie Ji[2], Le Zhang[4*], Peter Uhd Jepsen[2], Xin Yu[1], Shubin Yan[5], Huichang Li[1], Qian Shen[2], Daena Madhi[2], Binbin Zhou[2*], Xiaohua Deng[3*]

[1]School of physics and materials science, Nanchang University, Nanchang, 330031, China

[2]Department of Photonics Engineering, Technical University of Denmark, 2800 Kgs. Lyngby, Denmark

[3]Institute of Space Science and Technology, Nanchang University, Nanchang 330031, China

[4]Key Laboratory of Electromagnetic Wave Information Technology and Metrology of Zhejiang Province, College of Information Engineering, China Jiliang University, Hangzhou 310018, China

[5]College of electrical engineering, Zhejiang University of Water Resources and Electric Power, Hangzhou, 310018, China

*Corresponding authors. Email: shenyun@ncu.edu.cn; zhangle@cjlu.edu.cn; zhou@fotonik.dtu.dk; dengxhua@gmail.com

**Topological systems are inherently robust to disorder and continuous perturbations, resulting in dissipation-free edge transport of electrons in quantum solids, or reflectionless guiding of photons and phonons in classical wave systems characterized by topological invariants. Despite considerable efforts, direct experimental demonstration of theoretically predicted robust, lossless energy transport in topological insulators operating at terahertz frequencies is needed further investigations to shed affirmative light on the unique properties enabled by topological protection. Here, we introduce Kagome lattice that exhibits a new class of symmetry-protected topological phases with very low Berry curvature but nontrivial bulk polarization, and fabricate an optical topological insulator that provide the valley hall effect. Theoretical**

**analysis show that four type edge states can be obtained. Measurements of THz-TDs with high time-resolution demonstrate that terahertz wave propagating along the straight topological edge and Z-shape edge with sharp turns have almost same high transmission in 0.440 THz to 0.457 THz domain range. Those results quantitatively illustrate the suppression of backscattering due to the non-trivial topology of the structure. The THz-TDs measurement yields amplitude and phase information, showing significant advantage compared to general broadband infrared, single wavelength continuous-wave THz measurements and visible spectroscopy. It allows further exploration of the effective refractive index, group velocity and dispersion relations of edge states. Our work offers possibilities for advanced control of the propagation and manipulation of THz waves, and facilitates the applications including sixth-generation (6G) wireless communication, terahertz integrated circuits, and interconnects for intrachip and interchip communication.**

A topological insulator supports some of the most fascinating properties for signal transport and wave propagation. It insulates in the bulk, but conducts along the edge and offers unprecedented robustness to defects and disorder, advancing research fields from condensed matter physics to acoustics and photonics. Since the first photonic analogue of a quantum Hall topological insulator was realized in the microwave regime using gyromagnetic materials, in which a strong magnetic field was utilized to break the time-reversal symmetry, different kinds of photonic topological systems have been successively proposed and studied [1-11]. Among of those, non-magnetic quantum Hall topological insulators mimicking time-reversal-symmetry breaking were demonstrated at near-infrared frequencies using lattices of coupled waveguides and ring resonators. Photonic analogues of spin [12-19] and valley Hall effects [20-25] were demonstrated in the microwave and near-infrared frequency range. Whatever, most of the topological systems are characterized by topological indexes such as Chern-class numbers, Z2 invariants and winding numbers. More recently, a new class of symmetry-protected topological phases has been introduced theoretically, which has very low Berry curvature but nontrivial bulk polarization [28-31], offering an opportunity to implement robust, controllable physical response for acoustic and photonic devices based on topological bulk polarization. Particularly, at terahertz (THz) frequency, the topological transport based on graphene-like lattice by millimeter-wave signal generator was first studied by Y. Yang, et. al. in 2020 [32]. Nevertheless, theoretical prediction and experimental demonstration of robust, lossless energy transport in THz topological insulators operating is still needed to deeply explore.

Here, we design a Kagome lattice photonic crystal that exhibits a new class of symmetry-protected topological phases with low Berry curvature but nontrivial bulk polarization, and experimentally demonstrate photonic topological states in THz frequency range based on THz time-domain spectroscopy (THz-TDS) with high time-resolution [33]. Specifically, different type edge states are analyzed. Wave propagating along Z-shape topological edges with sharp turns is studied to directly illustrate the robust energy transport and quantitatively characterize the suppression of backscattering. Our results open new opportunities for topological photonic insulator and paves the way for direct THz-TDs measurement of topological photonic systems in free space.

**Results and discussion**

**Design and theory.** The proposed Kagome lattice photonic crystal is shown in Fig. 1a, which is composed of three identical air holes with $C_{3v}$ PGS (Point Group Symmetry) embedded in Si with permittivity $\varepsilon \approx 11.56$. The diameter of the air holes is $d=0.36a$, where $a$ is the lattice constant. Similar with the method in Ref [16], we explore the topological phase transition by shrinking or expanding the air-hole array within the unit cell. According to 2D Su-Schrieffer-Heeger model, the intracellular and intercellular coupling determine the topology of the system. Specifically, the dominant factors are severally the distance γ between two holes in same cell and γ' between two nearest holes in neighboring cells, or equivalently the center distance $W$ between each air hole and unit cell. One can find that when $W = a/(2\sqrt{3})$, the intracellular coupling equals to the intercellular coupling (|γ'|/|γ|=1). For this special case, the photonic band (TM modes with magnetic field along z axis) in red curves of Fig. 1b exhibit a Dirac-like degeneracy formed by the lowest two bands at K (K′) point in first Brillouin zone (BZ). When the distance $W$ shrinks ($W < a/(2\sqrt{3})$) or expands ($W > a/(2\sqrt{3})$), the intracellular and intercellular couplings are no longer equal, leading a photonic bandgap opened between the lowest two bands at K (K′) point, as shown the gray region in Fig. 1b. Especially, during the transition from $W < a/(2\sqrt{3})$ to $W > a/(2\sqrt{3})$, band inversion takes place between the 1st and 2nd bands at K (K′) degeneracy. For $W$=a/4 to $W$=a/3, the variations of field in band 1st, 2nd and 3rd are depicted in Fig. 1c, indicating the inversion.

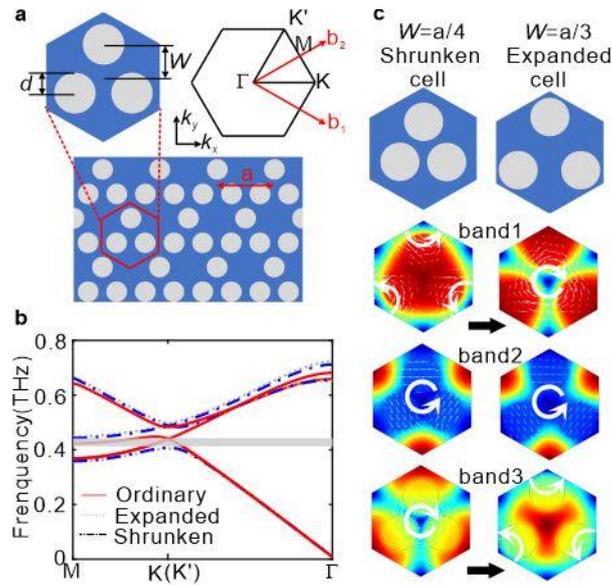

**Fig. 1. Kagome lattice photonic crystal and band inversion. a** Schematic of Kagome lattice photonic crystal. **b** Photonic band structure (TM modes with magnetic field along z axis). **c** Variations of field in bands 1st, 2nd, and 3rd when $W$=a/4 changes to $W$=a/3.

The topological phase transition can be demonstrated by topological theory.

Typically, the nth-band Berry curvature is defined as $F_{\mathbf{k}}^{(n)} = \nabla_{\mathbf{k}} \times A_{\mathbf{k}}^{(n)}$ with $A_{\mathbf{k}}^{(n)}$ denoting the Berry connection of the nth band [26]. For Kagome system, 2D bulk polarization, P=(p1, p2), with $p_i = -\frac{1}{(2\pi)^2} \iint_{BZ} dk_1 dk_2 \, \text{Tr}[A_i(\mathbf{k})]$ [27], can be used to characterize the topological invariant (see Supplementary). In which, $[A_i(\mathbf{k})]^{mn} = -i \langle u^m(\mathbf{k}) | \partial k_i | u^n(\mathbf{k}) \rangle$ is the Berry connection matrix with $m$ and $n$ run over occupied energy bands. $|u^n(\mathbf{k})\rangle$ is the periodic Bloch function for the nth band. $\mathbf{k} = k_1 \mathbf{b}_1 + k_2 \mathbf{b}_2$ is the wave vector. A direct integration shows that P=(1/3, 1/3) for $W > a/(2\sqrt{3})$ and P=(0, 0) for $W < a/(2\sqrt{3})$, indicating topological nontrivial and trivial phases, respectively (mathematical details see Supplemental Material). This polarization can also be obtained by making use of C3 PGS with $\exp(i2\pi p_{1,2}) = \prod_{n \in occ} \frac{\beta_n(K)}{\beta_n(K')}$, where $\beta_n(\mathbf{k})$ is eigenvalue of the operator R3 at C3-invariant k point on the nth band (see detail in Supplementary). Generally, topological edge states can appear at the interface separating two PCs with different bulk polarization.

**Edges states.** Next, we investigate the topological states between two PCs with different 2D polarization. In detail, as the band inversion takes place between the 1st and 2nd bands at K (K′) degeneracy of the Brillouin zone shown in Fig. 2a(i), the possible situations of edge states appearing at domain wall separated by the shrunken PC (W=a/4) and expanded PC (W=a/3) are demonstrated in Fig. 2a(ii) and (iii). For domain walls with 0° and ±120° turns in Fig. 2a(ii), the band inversions of edge states are at K degeneracy of Fig. 2a(i), and the surrounding configure is formed by shrunken PC with P=(0, 0) and expanded PC with P=(1/3, 1/3). For ±60° and 180° turns, it is K' degeneracy and by P=(0, 0), P=(-1/3, -1/3). Conversely, in Fig. 2a(iii), surrounding configures of K and K' are formed by P=(0, 0), P=(-1/3, -1/3) and P=(0, 0), P=(1/3, 1/3), respectively. To distinguish the situations, we severally mark the

K/K' in Figs. 2a (ii) and (iii) with K1/K1' and K2/K2'. Here, the normalized field ($|H_z|$) distribution, band diagram for situations of K1 and K1' are shown in Fig. 2b(i) and (ii), respectively.

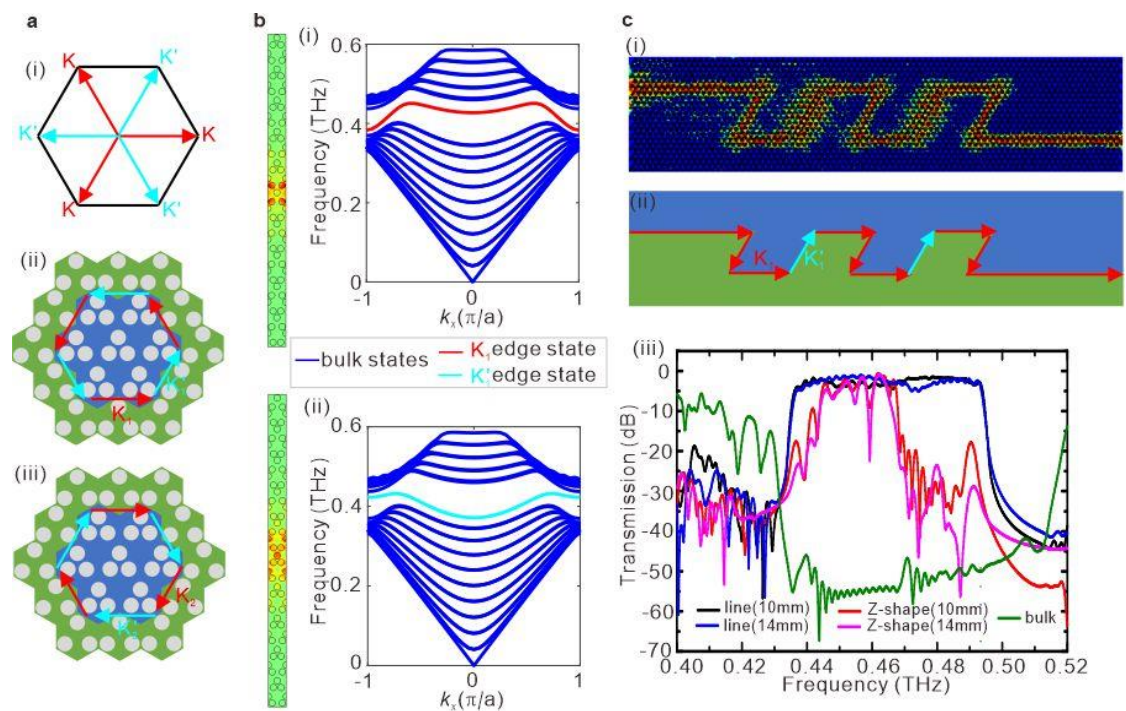

**Fig. 2 Properties of edge states. a** (i) Brillouin zone and (ii)-(iii) possible situations of edge states appearing at domain wall separated by the shrunken PC ($W=a/4$) and expanded PC ($W=a/3$). **b** Normalized field ($|Hz|$) distribution and band diagram for situations of (i) K1 and (ii) K1' edge states. **c** (i) Normalized field distribution of 14 mm Z-shape chip and (ii) its possible situations of edge states. **d** Transmissions of 10 mm and 14 mm chips with straight line, Z-shape domain walls, respectively.

As is known, one of the most intriguing properties of topological edge states is that they are immune to sharp corners scattering. To demonstrate this property, the straight-line, and highly twisted Z-shape with sharp corners domain wall on 10 mm, 14 mm length silicon chips are explored and performed by CST STUDIO SUITE (CST). We note that the 10 mm/14 mm Z-shape domain wall includes two/three 120° turns and one/two 60° turns. Particularly, the normalized field distribution of 14 mm Z-shape chip is illustrated in Fig. 2c(i) and Fig 2c(ii) demonstrates the situations of the edge states in the propagation. It is shown that K1 and K1' degeneracy both exist in the Z-shape chip. Comparatively, if it is a straight-line chip, there will be only K1.

The calculated transmissions of 10 mm/14 mm straight line, Z-shape domain walls and the bulk are demonstrated in Fig. 2c(iii). In which, distinct dip (green curve) between 0.43 THz and 0.51 THz can be clearly seen, indicating the presence of a bulk bandgap. In addition, transmissions of the two Z-shape domain walls (10mm and 14mm, red and pink curves) are almost equal in range 0.445 THz to 0.463 THz, while the two straight line domain walls (10 mm and 14 mm, black and blue curves) are almost equal in range 0.436 THz to 0.493 THz. Particularly, in range 0.445 THz to 0.463 THz, the transmissions for all domain wall chips are almost equal. That is, the length and the number of turns of domain wall have almost no effect on variation of transmissions in such frequency range, indicating the immunity of corners scattering and topological edge states properties. We note that in Fig. 2c(iii), transmission width of straight-line and twisted Z-shape are different. It is due to K1' and K1 have different transmission band (see Supplementary Information), while here K1, K1' degeneracy both exist in Z-shape and only K1 is in straight-line domain walls.

**Experiment results.** To experimental verify the above properties, straight line and highly twisted Z-shape domain walls samples on 8 ×10 mm2 and 8 ×14 mm2 (width × length) silicon chips are fabricated (see Supplementary Information), respectively. As an instance, microscope image of the 8 ×14 mm2 silicon chip is shown in Fig. 3a (i), which including three 120° and two 60° turns.

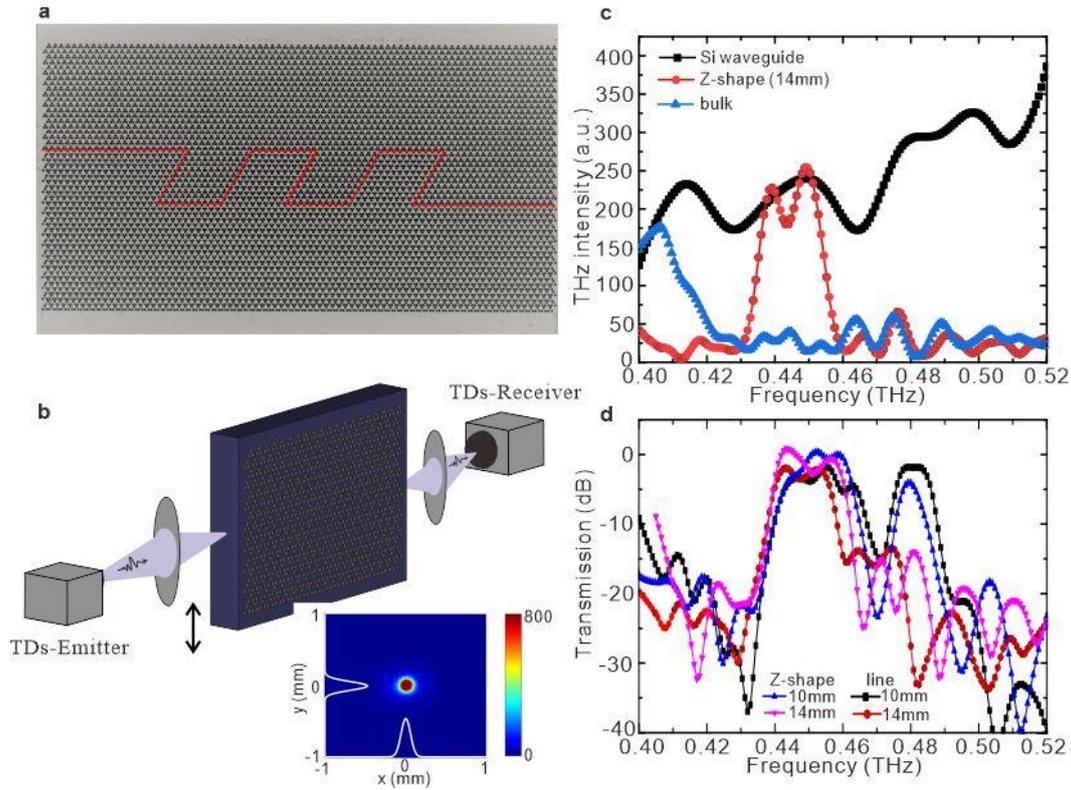

**Fig. 3 Experimental results. a** Microscope image of fabricated 8 ×14 mm$^2$ Z-shape twisted domain walls chip. **b** Schematic diagram of measurement setup. The spot size by THz camera is shown in the inset. **c** Measured transmission amplitudes of Si, bulk, and Z-shape domain wall of the chip in **a**. **d** Transmissions of the 10 mm and 14 mm fabricated chips with straight, Z-shape domain walls, respectively.

The transmission amplitude of topological states along the domain walls are measured by THz-TDs (see Methods and Supplementary Information) shown in Fig. 3b, where the spot size of THz beam on the chips measured by THz camera is shown in the inset. Figure 3c shows the transmission amplitudes of Si, bulk, and Z-shape domain wall in 8 ×14 mm2 sample chips. In which, a distinct dip in transmission, between 0.43 THz and 0.52 THz, can be seen for bulk of the chip, indicating the presence of a bulk bandgap. For the convenience of comparison, transmittance of the Z-shape 14 mm domain wall, obtained by the amplitudes ratio of domain wall to Si in Fig. 3c, is specially show in Fig. 3d. Successively, the transmittances of 10 mm straight-line and 10 mm Z-shape, 14mm Z-shape samples are measured and demonstrated in Fig. 3d, respectively. In which, transmittances of the 10 mm and 14mm Z-shape samples are almost equal in range 0.440 THz to 0.457 THz, while the two straight-line domain walls are almost equal in range 0.440 THz to 0.485 THz. That is, in range of 0.440 THz to 0.456 THz, the transmissions for all structure are almost equal. Those results agree well with the numerical prediction in Fig. 2c(iii) and confirm that the immunity of corners scattering and topological edge states properties. We note that in the experiment, the results are affected by many factors, for example, characters of Si including the dispersion and loss, the disturbance of residual Si at ends of the sample, and the resonance of the structure, which may lead to the transmission dip of straight-line near 0.47 THz and some deviations. Whatever, the transmittances in Fig. 3d are approximately equal in range of 0.440 THz to 0.456 THz indisputably verify the suppression of backscattering due to the non-trivial topology of the structure.

**Time resolution.** In the case of THz-TDs, the time-domain signal directly measures the transient electric field rather than its intensity. Specifically, the time-domain waveform of the electric field of the transmitted THz pulse is measured by scanning the optical delay stage. Then, spectral information is yielded through Fourier-transform. Figure 4a shows the THz-pulses transmitted through the fabricated straight-line, Z-shape domain walls of 10 mm, 14 mm chips, respectively. The corresponding frequency spectra are shown in Fig. 4b. In the Fourier-transform, we set 10 ps as a step to extract the time signal in sequence and transform it to frequency domain to accurately target the propagation process of THz wave and physical effects in the samples. It shows that the topological edge states' transmissions, which is in range 0.42 THz to 0.52 THz in Fig. 4b and correspond to same range in Fig. 3d, are mainly determined by the second pulse in Fig. 4a marked with dotted boxes. As the domain-wall lengths in the four fabricated chips are different, the centers of the second pulses are at different time positions. Here, the effective time domains for the topological edge states' transmissions in Fig. 4b (i) and (ii) are approximate 300 ps to

450 ps and 330 ps to 450 ps, respectively. They are broader than the domains in Fig. 4b(iii) and (iv), which are severally approximate 420 ps to 490 ps and 510 ps to 580 ps. We attribute such difference to the more explicit multi-reflection of Si margin area in straight line chips. In addition, we also can timely transform the whole-time domain to frequency domains in the process of time evolution, and dynamically present the variations of edge states' transmission (see Supplementary Information). Actually, the measurement yields amplitude and phase information is a significant advantage of THz-TDS compared to broadband infrared, single wavelength continuous-wave THz measurements, and visible spectroscopy. Measuring the spectra in this manner theoretically allows one to furtherly calculate the effective refractive index, group velocity and dispersion relations of edge states in Fig. 2b.

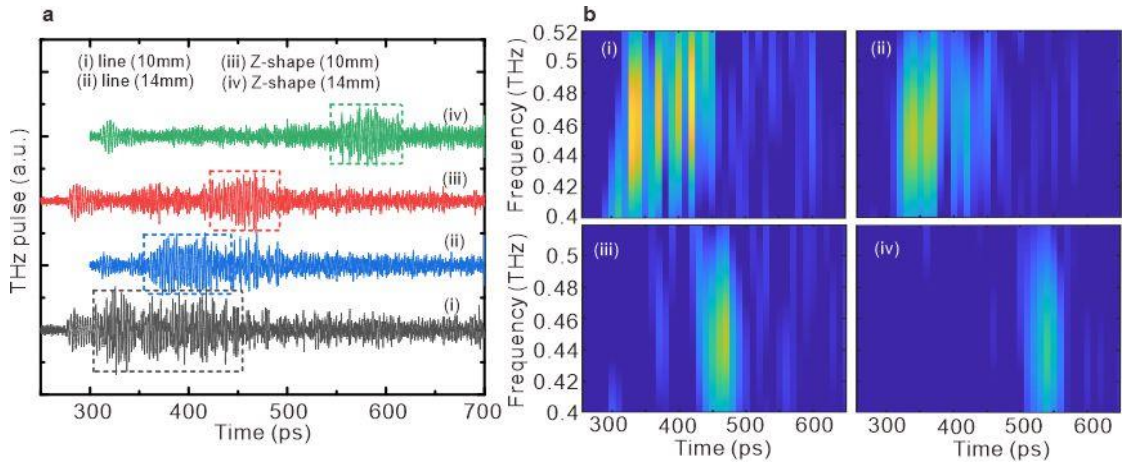

**Fig. 4 Time resolutions. a** Measured temporal profiles and **b** corresponding Fourier-transform frequency spectra of THz-TDs pulse through the fabricated (i)-(ii) straight line, (iii)-(iv) Z-shape domain walls of 10 mm, 14 mm chips, respectively.

In summary, we designed and experimentally demonstrated a silicon-based Kagome crystal structure that exhibits the valley Hall effect at THz wavelengths. Robust transport is observed by a direct and quantitative comparison of the light transmitted along the topological Z-shape domain walls with sharp turns and straight-line domain walls. Our study demonstrates the major fundamental phenomena of the new class of symmetry-protected topological phases, and extends the family of topological insulators from visible wavelengths to THz domain, paving the way for THz-TDs measurement in topological photonic systems and offering possibilities for advanced control of the propagation and manipulation of THz waves in applications including sixth-generation (6G) wireless communication, terahertz integrated circuits, and interconnects for intrachip and interchip communication.

**Methods**

**Samples fabrication.** In the fabrication, a standard 4-inch silicon wafer with a 1 µm thick silicon dioxide layer was prepared in two steps: (1) The top silicon dioxide layer was patterned with photoresist for ultraviolet (UV) lithography followed by the reactive ion etching (RIE); (2) silicon dioxide was used as a hard mask to etch 500 µm of silicon at the bottom by deep reactive ion etching (D-RIE) to finally get the straight and Z-shaped edge Kagome crystal samples (see Supplementary Information).

**Measurement.** The THz-TDs system (Toptica TeraFlash Smart) has two fiber-coupled photoconductive antennas which can generate and detect terahertz pulses with >4 THz bandwidth and over 90 dB dynamic range. In addition, the waist of the Gaussian-beam profile at the focal plane is determined with knife-edge method (see Supplementary Information).

**Acknowledgements**

This work is supported by the National Natural Science Foundation of China (Grant numbers 61865009, 61927813, 62165008).

**Author contributions**

Y.S. and J.J. have same contributions. Y.S. and L.Z. conceived the idea of this work. Y.S., L.Z., J.J., H.L., and X.Y. performed numerical simulation. Y.S., X.Y., and S.Y. fabricated the samples. Y.S., J.J., Q.S., B.Z., and D.M. performed the measurement. All authors contributed to writing the manuscript. P.J., B.Z., and X.D. supervised the project.

**Competing interests**

The authors declare no conflict of interests.

# Supplementary Information

**Berry curvature**

First, through calculation of the Berry curvature we show the traditional valley Chern number in the Kagome system is not a quantized topological invariant. Starting from the Maxwell's equations, we can obtain the band structure of the proposed Kagome-lattice photonic crystal (PhC). The *n*th-band Berry curvature is expressed as

$$F_{\mathbf{k}}^{(n)} = \nabla_{\mathbf{k}} \times A_{\mathbf{k}}^{(n)} = \nabla_{\mathbf{k}} \times \langle u_{\mathbf{k}}^{(n)} | \nabla_{\mathbf{k}} | u_{\mathbf{k}}^{(n)} \rangle \quad (1)$$

with $A_{\mathbf{k}}^{(n)}$ denoting the Berry connection of the *n*th band. $A_{\mathbf{k}}^{(n)}$ is gauge dependent but the Berry curvature is invariant under gauge transformation. Through an efficient algorithm in Ref. [1], we obtain the Berry curvature by calculating the Berry flux expressed in the Berry connection form of a closed Wilson loop in *k*-space,

$$F_k^{(n)} dS = \mathrm{Im}[\ln \langle u_k^{(n)} | u_{k+dk_1}^{(n)} \rangle \langle u_{k+dk_1}^{(n)} | u_{k+dk_1+dk_2}^{(n)} \rangle \langle u_{k+dk_1+dk_2}^{(n)} | u_{k+dk_2}^{(n)} \rangle \langle u_{k+dk_2}^{(n)} | u_k^{(n)} \rangle] \quad (2)$$

where *dS* is formed by **k**-space discretization with a moderate *m*-point mesh in the Brillouin Zone (BZ).

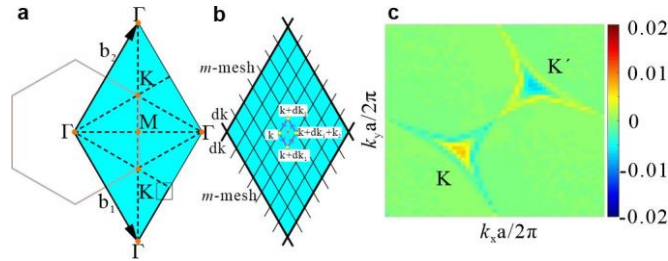

**FIG. S1 a** The Brillouin Zone and reciprocal lattice vectors of the hexagonal lattice. **b** The closed Wilson loop path under m×m *k*-space discretization. **c** Calculated Berry curvature of the Kagome lattice under a 50x50 mesh in *k*-space.

**Bulk polarizations**

By borrowing the theory of charge polarization in solids, the bulk polarization is used instead to characterize the topological phase in our Kagome-lattice PhC. In highly-symmetric photonic lattice, the bulk polarization can describe the Wannier center moving, which is related to the Zak phase given by the following expression

$$\mathbf{p} = \frac{1}{2\pi} \iint dk_1 dk_2 \, \text{Tr}[\mathbf{A}(k_1, k_2,)] \tag{4}$$

The polarization in our Kagome PhC is calculated using the Wilson loop approach. We assume wave vectors $k_1$ and $k_2$ have the directions respectively along the two reciprocal lattice vectors $b_1$ and $b_2$ in hexagonal cell shown in FIG.S1. Then the Wilson-loop along $k_1$ direction under $N$ mesh is defined as

$$W_{k_1 \to k_1+2\pi, k_2} = A_{k_1,k_2} A_{k_1+dk_1,k_2} A_{k_1+2dk_1,k_2} \cdots A_{k_1+Ndk_1,k_2} \tag{5}$$

where $A_{k_1,k_2} = \langle u_{k_1,k_2} | u_{k_1+dk_1,k_2} \rangle$ is the Berry connection for a short sector of the loop along $k_1$ direction, and $dk_1 = \frac{2\pi}{a(N+1)}$. The Wilson-loop along $k_2$ direction has a similar expression. Because the phases of the eigenvalues of the Wilson loop are Wannier centers, again using summation instead of integration under k-space discretization, we can write the polarization $\mathbf{p}=(p_1, p_2)$ as

$$p_1 = -\frac{i}{2\pi N} \sum_{k_2} \ln(W_{k_1 \to k_1+2\pi, k_2}), \quad p_2 = -\frac{i}{2\pi N} \sum_{k_1} \ln(W_{k_1,k_2 \to k_2+2\pi}) \tag{6}$$

In the shrunken Kagome lattice, the bulk polarization (0,0) shows its topologically trivial phase, indicating states are pinned to the center of the lattice. In the expanded Kagome lattice, the bulk polarization (1/3, 1/3) shows its topologically nontrivial phase, indicating states can be localized at the edges.

The polarization can also be demonstrated by making use of point group symmetry (PGS) theory. First, the connection under a rotation operator $\tilde{R}$ is expressed as

$$A_R(k) = \langle u_n(Rk) | \tilde{R} | u_n(k) \rangle \tag{6}$$

Since the Kagome lattice satisfies $C_3$ PGS, we can start from the integral of the Berry connection to obtain the polarization according to the method in [2]

$$p_j = \frac{1}{2\pi} \int dk_1 \int dk_2 \, \text{Tr}\left[A_j(k_1, k_2)\right] \tag{7}$$

Using the $C_3$ rotation operator, we have

$$p_j = \frac{i}{2\pi} \int dk_1 dk_2 \, \text{Tr}\langle u(C_3\mathbf{k}) | \frac{d}{dk_j} | u(C_3\mathbf{k}) \rangle + q_j(C_3) \tag{8}$$

where $q_1(C_3) = \frac{i}{2\pi} \int dk_2 \left( \int dk_1 \frac{d \ln\{\det[A_{C_3}(k_1, k_2)]\}}{dk_1} \right)$ (taking $j=1$ for example) is the winding number of $\det(A_{C_3})$ presenting a phase at $k_2$. Under $C_3$ PGS, the following constraints need to be satisfied

$$p_1 - p_2 = q_1(C_3), \quad p_1 + 2p_2 = q_2(C_3) \tag{9}$$

To calculate $q_1(C_3)$, the phase of the determinant of the rotational connection matrix associated with $C_3$ is investigated in the BZ

$$\phi(\mathbf{k}) = -i \ln \det(A_{C_3})(\mathbf{k}) \tag{10}$$

At specific K/K' points, $\phi(K)$ and $\phi(K')$ can be determined by the $C_3$ eigenvalues

$$\phi(K) = -i \ln \left[\prod_{n \in occ} \theta_n(K)\right], \quad \phi(K') = -i \ln \left[\prod_{n \in occ} \theta_n(K')\right] \tag{11}$$

By expressing $\phi(\mathbf{k})$ at six corners of the BZ in the form of $\phi(K), \phi(K')$ and $q_j(C_3)$, the following equation is derived

$$\phi(K') + 2q_2\pi - \phi(K) = 2q_1\pi - 2[\phi(K') - \phi(K)] \tag{12}$$

Thus

$$q_1(C_3) - q_2(C_3) = \frac{3}{2\pi}\left[\phi(K') - \phi(K)\right] \qquad (13)$$

Combining the constraints in Eq. (9), the final expression of the polarization is

$$\exp(i2\pi p_{1,2}) = \prod_{n \in occ} \frac{\theta_n(K)}{\theta_n(K')} \qquad (14)$$

With Eq. (14), we can get the same polarization (0,0) in the shrunken Kagome lattice and the same polarization $(\frac{1}{3}, \frac{1}{3})$ in the expanded Kagome lattice.

**Topological edge modes Simulations**

For domain walls with 0° and ±120° turns in Fig. 2a(ii), the band inversions of edge states are at K degeneracy of Fig. 2a(i). The surrounding configure is formed by shrunken PC with P=(0, 0) and expanded PC with P=(1/3, 1/3). For ±60° and 180° turns, it is K' degeneracy and by P=(0, 0), P=(-1/3, -1/3). The normalized field (|Hz|) distributions for situations of K1-K1-K1 and K1-K1'-K1 are shown in Fig. S2a and b, respectively. The corresponding transmissions are shown in Fig. S2c. From c we can see, the range of the transmissions are different.

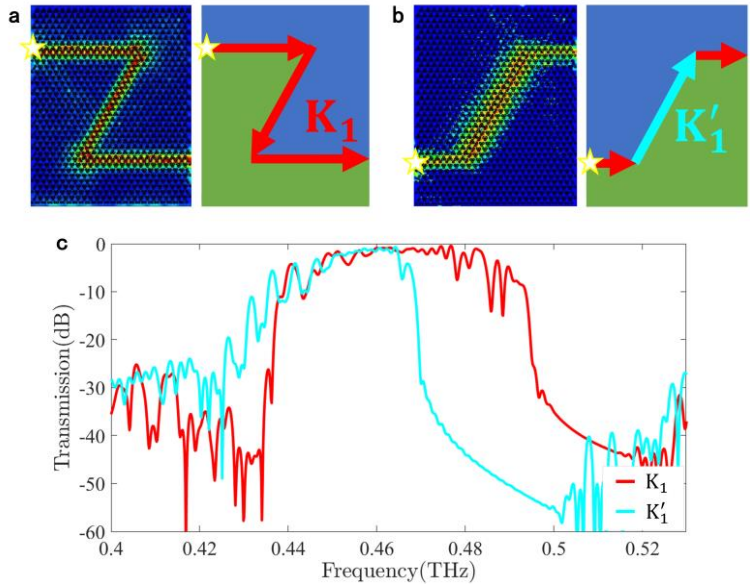

**Fig. S2** The normalized field (Hz) distributions for situations of **a** K1-K1-K1 and **b** K1-K1'-K1. **c** Corresponding transmissions of **a** and **b**.

**Fabrication.**

A standard 4-inch silicon wafer with 1µm thick silicon dioxide layer is prepared. The micro-fabrication steps for the samples are schematically shown in Fig. S3a-f.

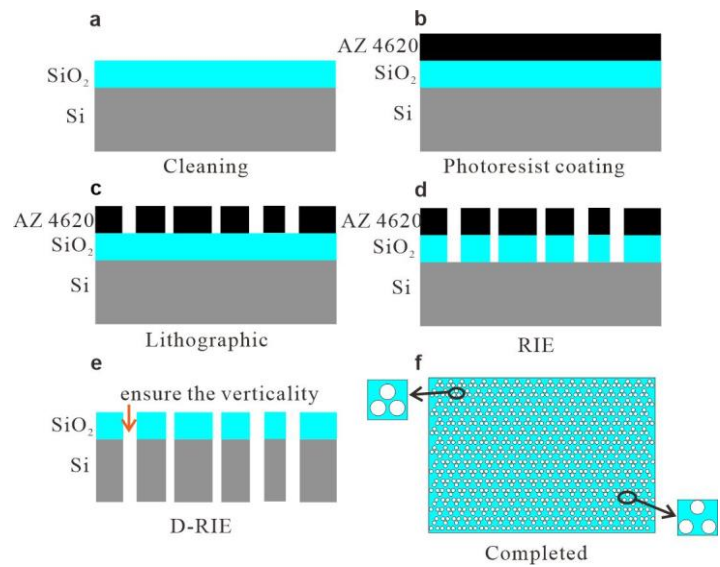

**Fig. S3 a-f** Micro-fabrication steps of samples.

In the step shown in Fig. S3a, the silicon wafer is cleaned and dried according to the (RCA) standard to remove organic or inorganic contaminants. Then, spin coating with a thin layer of hexamethyldisilazane (HMDS) primer is employed. This step will ensure a strong adhesion of the photoresist to the substrate groups[1].

In the step shown in Fig. S3b, a photoresist (AZ 4620) of constant thickness is spin-coated onto the front of the silicon wafer. Afterward, the sample is placed for 10 minutes in an oven to remove any excess solvent remaining in the photoresist and to reduce the stress inside the resist to increase the adhesion of the photoresist layer to the wafer.

In the step shown in Fig. S3c, the silicon wafer is photolithographically patterned via UV light exposure through a glass plate as a photo mask. After the exposure, the silicon wafer is immersed into the positive developer AZ4620 and shaken until the pattern is clearly visible. The sample is rinsed in water and dried. Then, it is placed in an oven for 30 minutes to remove the residual developer and anneal the photoresist layer, thereby enhancing the interfacial adhesion between the photoresist and wafer and further increasing the hardness of the photoresist, which is preferable for dry etching.

In the step shown in Fig S3d, the pattern was transferred to Silica by RIE in $C_4F_8$ and Ar gases atmosphere, performed by a plasma etch system. Thoroughly clean the residue after the RIE process, as this is essential for the second manufacturing step. We use NMP remover heated on a hot plate for 40 minutes to 30 minutes, and then rinse with acetone and ethanol. Finally, rinse with nitrogen in deionized water and blow dry.

In the step shown in Fig. S3e, use silicon dioxide as a hard mask to etch 500μm of silicon at the bottom. It's a powerful maskless method for nanowireformation. We used patterned silicon dioxide as a mask for the second etch process. The pattern was transferred to silicon by D-RIE in $C_4F_8$ and $SF_6$ gases atmosphere, performed by an Omega LPX Dsi plasma etch system. In this step, it is important to ensure the verticality of the etching contour. Non-vertical etched contours induce coupling between polarization modes like TE and TM, thereby significantly increasing the loss in propagation. In addition, the etched sidewalls must be smooth to prevent scattering. The side walls of the prefabricated structure are almost vertical and smooth, ensuring low external losses.

In the step shown in Fig. S3f, removal of surface silica Soak in FH for 3 minutes, then rinse in deionized water, dry the wafer, and get the samples.

The microscope image of the sample is shown in Fig. 3a.

**Gaussian beam-profiling**

Standard knife-edge (KE) method is also employed to characterize the profile of THz beam. A metallic blade translates perpendicular to the optical axis. The intensity of unmasked portion and blade position are recorded to reconstruct the field profile. At location of Gaussian-beam waist, electric field for 0.8 THz resolved from Fourier transform versus the blade position is shown in Fig. S4. Here, the 0.8 THz is the centre frequency of Gaussian pulse in Toptica TeraFlash system. By performing a spatial derivation of the recorded signal in Fig. S4, the normalized field distribution is illustrated in Fig. S5 by red points, while pattern of Gaussian fit is shown with black curve. In which the FWHM of the electric field is 450 µm.

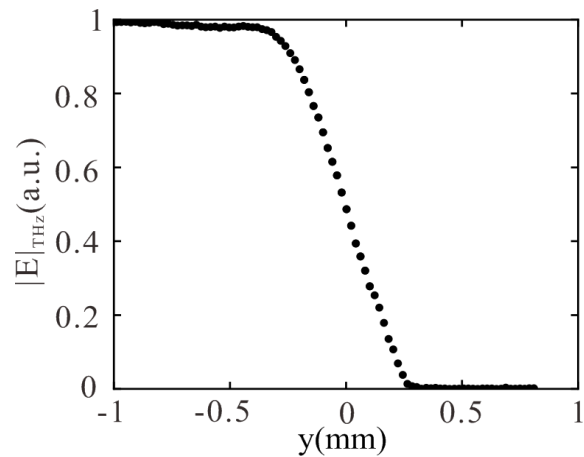

**Fig. S4** THz electric field at 0.8THz versus blade position.

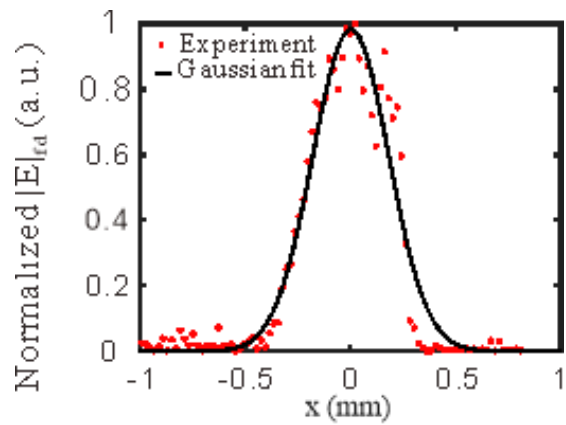

**Fig. S5** Gaussian beam-profiling.

The results agree well with that from THz-camera.

**Measurement process.**

Out THz time-domain spectroscopy system (Toptica TeraFlash Smart) has two fiber-coupled photoconductive antennas which can generate and detect terahertz pulses with >4 THz bandwidth and over 90 dB dynamic range. In the measurement, we take following steps to prepare the setup in Fig. 3b. Firstly, we keep the THz-emitter fixed and move the THz-receiver along the same height level to find the maximum pulse peak amplitude. In this situation, the THz beam is focused on onto a place in free space and then imaged onto the receiver through its focusing optics. Then, the focusing point of emitter in the free space will be used as the input port in the subsequent transmission measurements and therefore it needs to be located. Thus, a thin metal slab with 1.5 mm aperture hole, was aligned along propagation direction to find the location of highest transmission, corresponding to the focal plane of the emitter. We then repeated the procedure with a 600-µm aperture.

Secondly, we mount sample in its holder and place the input face of the waveguide in the focal plane. We then shift the position of the receiver corresponding to the sample width so the output facet of the waveguide is in the focal plane of the receiver.

In this manner we can shift the sample up and down and determine the transmission spectra (see details in the supplementary). We record 800 time-domain traces per second, and average 100 scans before calculating the transmission.

When the setup in Fig. 3b is prepared for the straight edge waveguide sample which is clipped with a holder on 3-axis stage, THz-receiver and THz-emitter are at same height level. Then we move the sample up and down near the edge interface by adjusting the stage to scan the transmission spectra, and find the maximum transmission in bandgap of 0.45 to 0.55 THz, which indicates that the light exactly propagate along the edge.

The transmission determining for Z-shaped sample is similar to the way for straight edge sample, except that in measurement of Z edge propagating, the receiver and emitter need has an approximate 1.5 mm height difference.

**Time resolution**

In fact, we also can in the process of time evolution, timely transform the whole-time domain to frequency domains, and dynamically present the variations of edge states' transmission. The time-domain signal of 14mm Z-shape sample in the process of evolution is shown in Fig. S6a. The corresponding Fourier-transform frequency spectra are shown in Fig. S6b.

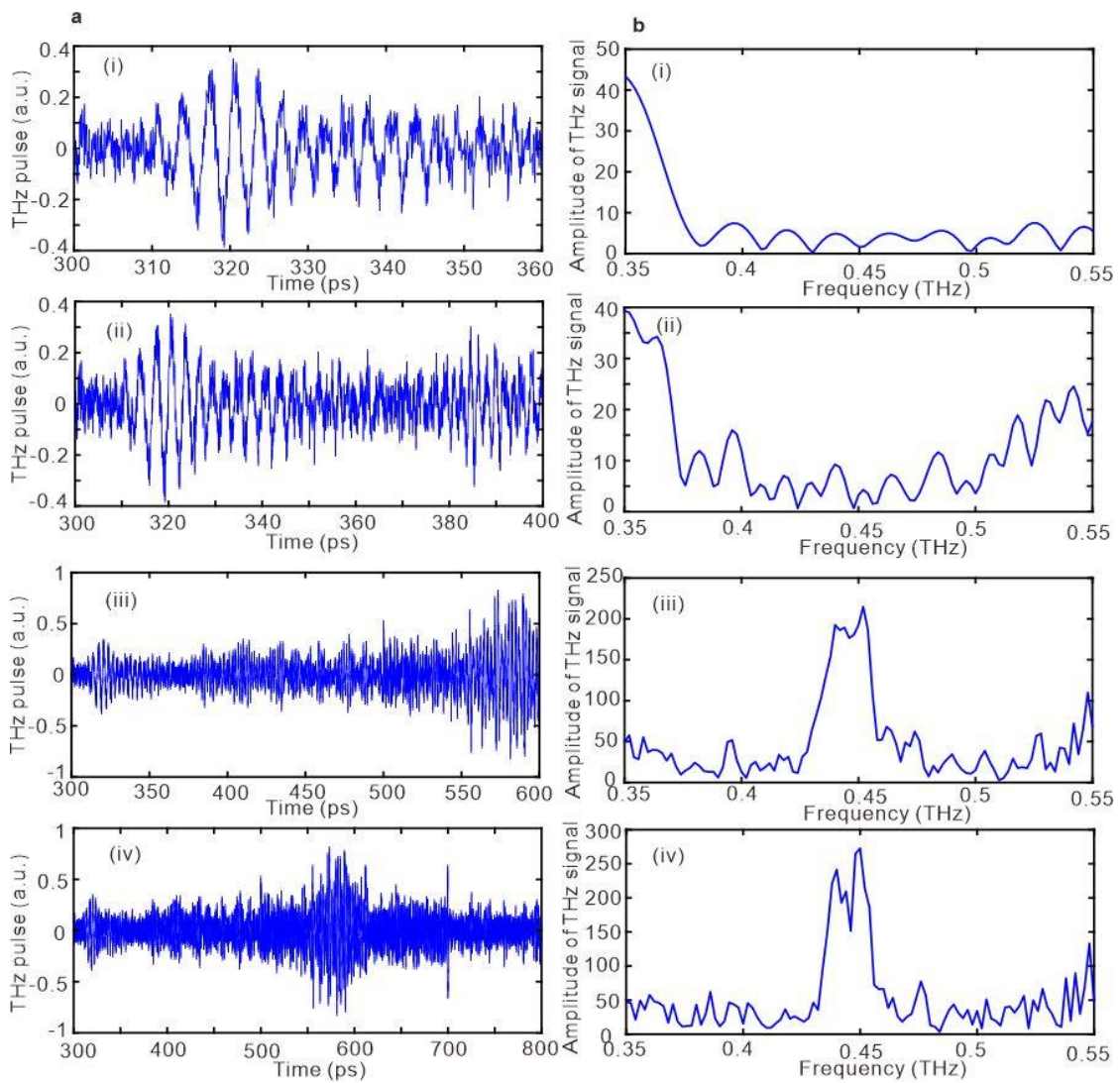

**Fig. S6 a** Measured temporal profiles and **b** corresponding Fourier-transform frequency spectra of 14mm Z-shape chip in the process of time evolution.